\documentclass[paper,11pt]{article}
\pdfoutput=1

\usepackage[pdftex,bookmarks,colorlinks]{hyperref}
\usepackage[pdftex]{graphicx}
\usepackage[amssymb]{SIunits}
\usepackage{mathrsfs,mathcomp}
\usepackage{fancyhdr}

\usepackage{hyperref}

\fancypagestyle{plain}{
\fancyhead[CO]{\footnotesize{Fluid dynamics video for the 31st Annual Gallery of Fluid Motion\\ 66th Annual Meeting of the American Physical Society, Division of Fluid Dynamics \\Pittsburgh, PA, Nov 2013. Entry \#102350}}
}

\begin{document}

\title{Fast and Scalded: Capillary Leidenfrost Droplets in micro-Ratchets}

\author{\'Alvaro Mar\'in$^{1}$\footnote{alvarogum@gmail.com}, Daniel del Cerro$^{2}$, Gert-Willem R\"omer$^{2}$, Detlef Lohse$^{3}$\\
\emph{$^1$Institute of Fluid Mechanics, Universit\"{a}t der Bundeswehr M\"{u}nchen, Germany}\\
\emph{$^2$ Applied Laser Technology, University of Twente, The Netherlands}\\
\emph{$^3$ Physics of Fluids Group, University of Twente, The Netherlands}\\}

\date{}
\maketitle

In the Fluid Dynamics Video included in the ancillary files (a different version is also available on \url{http://youtu.be/CS0c05WQ_js}), we illustrate the special dynamics of Capillary self-propelled Leidenfrost droplets \cite{Linke:Leidenfrost}\cite{lagubeau2011leidenfrost} and confirm the so-called ``viscous mechanism'' model \cite{Dupeux:2011} by testing it in micrometric ratchets with capillary droplets. In order to be able to propel water droplets of sizes of the order of 1 mm, micro-ratchets were produced by direct material removal using a picosecond pulsed laser source. Surface micro-patterning with picosecond laser pulses allows creating a well controlled topography on a variety of substrates, with a resolution typically in the micron range\cite{arnaldo2010}. The experiments yielded the surprising result that capillary drops can be much faster, and be propelled as much as bigger droplets. Based on the viscous mechanism model by D. Qu\'er\'e and C. Clanet \cite{Dupeux:2011} and adapting their scaling laws to capillary drops we obtain good agreement with the experimental results. More information can be found in reference \cite{gomez2012capillary} and \cite{marin2012capillary}.


\bibliographystyle{unsrt}
\bibliography{bib-ratchet}

\begin{thebibliography}{1}

\bibitem{Linke:Leidenfrost}
H.~Linke, B.~Alem{\'a}n, L.~Melling, M.~Taormina, M.~Francis, C.~Dow-Hygelund,
  V.~Narayanan, R.~Taylor, and A.~Stout.
\newblock {Self-Propelled Leidenfrost Droplets}.
\newblock {\em Physical Review Letters}, 96(15), April 2006.

\bibitem{lagubeau2011leidenfrost}
G.~Lagubeau, M.~Le~Merrer, C.~Clanet, and D.~Qu{\'e}r{\'e}.
\newblock Leidenfrost on a ratchet.
\newblock {\em Nature Physics}, 7(5):395--398, 2011.

\bibitem{Dupeux:2011}
G.~Dupeux, M.~Le~Merrer, G.~Lagubeau, C.~Clanet, S.~Hardt, and
  D.~Qu{\'e}r{\'e}.
\newblock {Viscous mechanism for Leidenfrost propulsion on a ratchet}.
\newblock {\em Europhysics Letters}, 96:1--7, November 2011.

\bibitem{arnaldo2010}
D.~Arnaldo~del Cerro, G.~R{\"o}mer, and A.~J. Huis In't~Veld.
\newblock Erosion resistant anti-ice surfaces generated by ultra short laser
  pulses.
\newblock {\em Physics Procedia}, 5:231--235, 2010.

\bibitem{gomez2012capillary}
A.~G. Marin, Arnaldo del Cerro, D., G.~W. R\"omer, B.~Pathiraj, A.~Huis in~'t
  Veld, and D.~Lohse.
\newblock Capillary droplets on leidenfrost micro-ratchets.
\newblock {\em Physics of fluids}, 24(12):1--10, 2012.

\bibitem{marin2012capillary}
A.~G. Marin, Arnaldo del Cerro, D., G.~W. R\"omer, B.~Pathiraj, A.~Huis in~'t
  Veld, and D.~Lohse.
\newblock Capillary droplets on leidenfrost micro-ratchets.
\newblock {\em arXiv preprint arXiv:1210.4978}, 2012.

\end{thebibliography}

\end{document}